\documentclass[twocolumn,trackchanges]{aastex63}

\usepackage{threeparttable}
\usepackage{epstopdf}
\usepackage{tabularx}
\usepackage{afterpage}
\usepackage{natbib}
\setcitestyle{notesep={; }}

\begin{document}

\title{WIYN$^5$ Open Cluster Study LXXIX. M48 (NGC 2548) I. Radial Velocities, Rotational Velocities, and Metallicities of Stars in the Open Cluster M48 (NGC 2548)}

\author{Qinghui Sun $^4$}
\author{Constantine P. Deliyannis $^4$}
\affil{Department of Astronomy, Indiana University, Bloomington, IN 47405, USA \\ qingsun@indiana.edu; cdeliyan@indiana.edu}
\footnote[4]{Visiting Astronomer, Kitt Peak National Observatory, National Optical Astronomy Observatory, which is operated by the Association of Universities for Research in Astronomy (AURA) under cooperative agreement with the National Science Foundation.}
\author{Aaron Steinhauer}
\affil{Department of Physics and Astronomy, State University of New York, Geneso, NY 14454, USA \\ steinhau@geneseo.edu}
\author{Bruce A. Twarog}
\author{Barbara J. Anthony-Twarog}
\affil{Department of Physics and Astronomy, University of Kansas, Lawrence, KS 660045, USA \\ btwarog@ku.edu; bjat@ku.edu}

\footnote[5]{The WIYN Observatory is a joint facility of the University of Wisconsin-Madison, Indiana University, the National Optical Astronomy Observatory and the University of Missouri.}

\begin{abstract}

WIYN/Hydra spectra (R $\sim$ 13,500, signal-to-noise pixel$^{-1}$ = 50--1000) of a 400 \AA\ region around Li 6708 \AA\ are used to determine radial and rotational velocities for 287 photometrically selected candidate members of the open cluster M48. The sample ranges from turnoff A stars to late-K dwarfs and eight giants. We combine our $V_{\rm{RAD}}$ measurements and power spectrum analysis with parallax and proper motion data from Gaia DR2 to evaluate membership and multiplicity. We classify 152 stars as single cluster members, 11 as binary members, 16 as members of uncertain multiplicity, 56 as single nonmembers, 28 as single ``likely" nonmembers, two as single ``likely" members, one as a binary ``likely" member, five as binary nonmembers, 10 as ``likely" members of uncertain multiplicity, three as nonmembers of uncertain multiplicity, and three as ``likely" nonmembers of uncertain multiplicity. From a subsample of 95 single members, we derive $V_{\rm{RAD}}$ = 8.512 $\pm$ 0.087 km s$^{-1}$ ($\sigma_{\mu}$, and $\sigma$ = 0.848 km s$^{-1}$). Using 16 isolated Fe I lines for a subsample of 99 single members (that have $\sigma_{T_{\rm{eff}}}$ $<$ 75 K (from 10 colors from $UBVRI$), {\it v} sin {\it i} $<$ 25 km s$^{-1}$, and well-behaved Fe I lines), [Fe/H]$_{\rm{M48}}$ = -0.063 $\pm$ 0.007 dex ($\sigma_{\mu}$). [Fe/H] is independent of $T_{\rm{eff}}$ over an unprecedentedly large range of 2500 K. The minimum cluster binary fraction is 11\%--21\%. M48 exhibits a clear but modest broadening of the main-sequence turnoff, and there is no correlation between color and {\it v} sin {\it i}.

\end{abstract}


\section{Introduction} \label{sec:intro}

M48 (NGC 2548; $\alpha$2000 = 8$^h$13$^m$43$^s$, $\delta$2000 = −5$\degr45\arcmin$) is a moderately rich, nearby ($D$ = 729 $\pm$ 26 pc), low-reddening ($E$($B-V$) = 0.05 $\pm$ 0.01 mag) open cluster with age (age = 420 $\pm$ 30 Myr, Deliyannis et al.\citet[in preparation, Paper II]{Deliyannis20}) intermediate to that of the Pleiades ($\sim$100 Myr) and the Hyades ($\sim$650 Myr). These characteristics make it a very interesting target for studying the rotational evolution and lithium (Li) depletion of stars, among many other topics. Of particular importance for nearby clusters younger than the Hyades is the ability to separate the often poor-to-moderately populated main-sequence cooler than the sun from the rising tide of field stars at fainter magnitudes encompassed by the large areal coverage of a nearby cluster, making evolutionary studies of lower-mass stars as a function of age a challenge. Equally critical for discerning any underlying link between fundamental stellar parameters (e.g. $T_{\rm{eff}}$, mass, {\it v} sin {\it i} and evolutionary state, as defined by position within the color-magnitude diagram (CMD), and internal evolution, as defined by atmospheric abundance changes) is the ability to separate single stars from binaries. As an example for Li studies, among the mechanisms proposed to create the severe F-dwarf lithium depletion\citet[the ``Li Dip,'' Boesgaard \& Tripicco][]{Boesgaard86}, mass loss and diffusion act closer to the age of the Hyades, whereas rotational mixing acts closer to the age of the Pleiades (Deliyannis et al.\citet{Deliyannis98}, Cummings et al.\citet{Cummings17}), so M48 should help delineate the evolution of the Li Dip and may help distinguish between proposed mechanisms. As another example, M48 can help delineate the post-Pleiades main-sequence depletion of Li in G dwarfs, which requires a mechanism(s) beyond the realm of ``standard'' theory (Deliyannis et al.\citet{Deliyannis90}, Cummings et al.\citet{Cummings17}). Finally, the age of M48 provides an important link in understanding the spindown of main-sequence stars.

Following a few early studies (Ebbighausen\citet{Ebbighausen39}; Li\citet{Li54}), and excepting studies limited to bright stars\citet[e.g. Baumgardt et al.][]{Baumgardt00}, the only modern proper motion study of M48 was that of Wu et al.\citet{Wu02}. With the evolution to Gaia DR2\citet{Gaia16, Gaia18}, this aspect of the cluster's database has changed dramatically, a point we will return to in Section \ref{sec:rv}. Spectroscopically, radial velocity studies have been restricted to the cluster's few giants or brightest main-sequence stars (Wallerstein et al.\citet{Wallerstein63}, Geyer \& Nelles\citet{Geyer85}, Mermilliod et al.\citet{Mermilliod08a}). Spectroscopic abundance analysis has been limited to one giant\citet[Wallerstein \& Conti][]{Wallerstein64}.

Photolelectric photometry was published by Pesch\citet[$UBV$, 37 upper main-sequence and giant stars]{Pesch61} and Claria\citet[DDO, five giants]{Claria85}. CCD photometry of thousands of stars in the direction of M48 has been reported in Wu et al.\citet[BATC 13-color]{Wu05}, Rider et al.\citet[$u'g'r'i'z'$]{Rider04}, Balaguer-Nunez et al.\citet[ubvy-$H\beta$]{Balaguer05}, and Paper II.

The present study is the first in a series of studies of M48. Here, we report radial velocities and {\it v} sin {\it i} for nearly 300 candidate members of M48. Together with Gaia DR2 data, we evaluate membership for each star, separate single stars from binaries/multiples, and discuss the binary fraction of the cluster. Finally, we conduct the first detailed spectroscopic metallicity of the cluster and discuss the result in the context of properties of open clusters in the solar neighborhood. Paper II presents $UBVRI$ photometry in the direction of M48 and reevaluates the basic cluster parameters. Paper III\citet[C.P. Deliyannis et al.][in preparation]{Deliyannis20b} presents Li abundances in M48 giants and from the turnoff to K dwarfs and addresses physical mechanisms that act to alter the surface Li abundances of stars.

\section{Observations and Data Reductions} \label{sec:obs}

Observations of M48 candidate members were made using the WIYN 3.5m telescope and Hydra multi-fiber spectrograph during four runs in 2017 October, 2017 December, 2018 March, and 2018 April. We used the 316@63.4 echelle grating in order 8 with the X19 filter, the blue cable, and the STA1 detector. The spectra span 6450--6850 \AA\ and have a dispersion of 0.205 \AA pixel$^{-1}$, and a resolution of R $\sim $13,500 as determined from our arc spectra (below).

Candidates for observation with WIYN/Hydra were chosen from our $UBVRI$ photometry (Paper II) as follows. For stars with $\bv\leq$ 0.40 mag, all stars on the by-eye photometric sequence were kept, including those off of the single-star sequence that might include rapid rotators and binaries. For $\bv>$ 0.40 mag, where galactic contamination increases, only stars on the single-star fiducial sequence with an approximate width of 0.02 mag in $\bv$, were kept; all five filters were used in defining this fiducial, which helps increase the fraction of members (see Paper II). In total, 287 stars were observed with Hydra using seven distinct configurations, which were made based on the $V$ magnitude and position of the star on the CMDs. For each configuration, Table \ref{tab:config} shows the configuration's name, approximate $V$ and $\bv$ ranges for most stars in the configuration, and the number of stars observed.

To help minimize errors, for each configuration, the following calibrations were taken in the same configuration as the object spectra: multiple Th Ar lamp spectra (both long and short), at least 11 dome flats, and daytime sky spectra (except for m48rg). Table \ref{tab:obs} provides the nightly log of observations. In total, we observed 4hr for m48vb1, 5.2hr for m48vb2, 8hr for m48b, 14.8hr for m48m1, 9.7hr for m48m2, 14.5hr for m48f, and 0.17hr for m48rg.

For each configuration, the raw spectra were bias-subtracted, flat-fielded, daytime sky spectra corrected, and wavelength-calibrated using IRAF. For the radial velocity ($V_{\rm{RAD}}$) work, cosmic-rays were eliminated with L.A. Cosmic\citet[van Dokkum][]{van01}. We first combined all of the reduced spectra of the same configuration for each night separately. For those configurations observed on more than one night, we shifted the night's average wavelength to match the cluster average, and then combined the spectra from different nights; for details and final membership and multiplicity results, see Section \ref{sec:rv}. We then normalized the combined spectra for each single member by fitting an eighth-order polynomial to the continuum, and used the normalized spectra to measure equivalent widths of iron lines to determine the stellar and cluster metallicities.

\begin{deluxetable}{ccccc}
	\label{tab:config}
	\tablecaption{Hydra Configurations}
	\tablewidth{700pt}
	\tabletypesize{\scriptsize}
	\tablehead{
		\colhead{Description} &
		\colhead{Name} &
		\colhead{$V$ range (mag)} &
		\colhead{$\bv$ range (mag)} &
		\colhead{\# Stars$^1$}
	} 
	\startdata
	\hline
	red giant & m48rg & 8.108 & 1.233 & 1 \\ 
	very bright 1 & m48vb1 & 9.119--11.671 & 0.029--1.313 & 55 \\
	very bright 2 & m48vb2 & 10.986--13.960 & 0.105--0.599 & 59 \\
	bright & m48b & 10.958--15.184 & 0.121--0.808 & 54 \\ 
	medium 1 & m48m1 & 15.042--17.059 & 0.768--1.226 & 47 \\
	medium 2 & m48m2 & 14.439--15.925 & 0.668--0.981 & 23 \\
	faint & m48f & 16.015--17.392 & 0.982--1.311 & 53 \\
	\hline
	\enddata
	\tablecomments{1. Configurations m48b and m48m2 both included star 2213; configurations m48vb2 and m48b both included star 2157; configurations m48m1 and m48m2 both included stars 2210, 2212, and 2221.}
\end{deluxetable}

\begin{deluxetable*}{lllrrrrrrlll}
	\label{tab:obs}
	\tablecaption{M48 observing logs}
	\tablewidth{700pt}
	\tabletypesize{\scriptsize}
	\tablehead{
		\colhead{Nights $^1$} &
		\colhead{Configurations} &
		\colhead{Exposure Time $^2$} &
		\colhead{Standards $^3$} &
		\colhead{$ < 1\sigma\ ^4$} &
		\colhead{1--2$\sigma$} & 
		\colhead{$ > 2\sigma$} & 
		\colhead{$V_{\rm{RAD}}$(km s$^{-1}$)$^5$} &
		\colhead{$\sigma_{V_{\rm{RAD}}}$ (km s$^{-1}$) $^5$}
	} 
	\startdata
			\hline
			1710n4 = 2017 Oct 31 & m48vb2 & 2hr & yes & 3 & 1 & 0 & 8.36 & 2.07\\ 
			1710n5 = 2017 Nov 1 & m48vb1, m48rg & 1.5hr, 10 minutes & yes & 2 & 0 & 0 & 12.37 & 6.73 \\ 
			1712n1 = 2017 Dec 21 & m48f & 3.33hr & yes & 0 & 1 & 5 & 7.93 & 5.54 \\ 
			1712n2 = 2017 Dec 22 & m48f & 3.92hr & yes & 1 & 1 & 1 & 8.52 & 1.20 \\
			1712n3 = 2017 Dec 23 & m48f & 7.33hr & yes & 0 & 2 & 2 & 8.53 & 0.31 \\ 
			1712n4 = 2017 Dec 24 & m48m1 & 7.33hr & yes & 1 & 0 & 0 & 8.52 & 1.29 \\
			1712n5 = 2017 Dec 27 & m48b & 0.92hr & no & 0 & 0 & 0 & 8.11 & 1.39 \\ 
			1712n6 = 2017 Dec 28 & m48b & 7.17hr & yes & 1 & 0 & 2 & 8.52 & 1.41 \\ 
			1712n7 = 2017 Dec 29 & m48vb1, m48vb2 & 2.5hr, 3.17hr & yes & 0 & 1 & 2 & 10.76, 8.46 & 4.97, 2.01 \\
			1803n1 = 2018 Mar 12 & m48m2 & 40 minutes & yes & 4 & 0 & 0 & 8.41 & 1.49 \\
			1803n2 = 2018 Mar 13 & m48m2 & 5.5hr & yes & 4 & 0 & 0 & 8.43 & 0.78 \\
			1804n1 = 2018 Apr 8 & m48m1 & 4hr & yes & 2 & 1 & 1 & 8.45 & 1.03 \\
			1804n2 = 2018 Apr 9 & m48m2 & 3.5hr & yes & 3 & 1 & 0 & 8.09 & 1.62 \\
			1804n3 = 2018 Apr 10 & m48m1 & 3.5hr & yes & 1 & 1 & 0 & 8.30 & 1.04 \\
			\hline
		\enddata
	\tablecomments{1. Dates of the observation, e.g., 1710n4 means the data were taken on the fourth night of the oserving run that began during 2017 October observing run, and the UT date is 2017 October 31. Afternoon calibrations may have begun on the previous UT date.
		2. The total exposure for the given configuration(s).
		3. Whether radial velocity standards were observed during the night.
		4. The number of radial velocity standards that fall within $1\sigma$, between $1\sigma$ and $2\sigma$, and above $2\sigma$ compared to the literature.
		5. Average radial velocity of each configuration determined by fitting a Gaussian profile to all of the observed stars of that configuration, and 1$\sigma$ error of the Gaussian fit.}
\end{deluxetable*}

\section{Radial Velocity, Binarity, Cluster Membership, and Cluster Binary Fraction} \label{sec:rv}

To determine a cluster average $V_{\rm{RAD}}$ and metallicity, we used a suitably constrained subset of cluster member single stars. The following subsections describe how we determined the multiplicity and membership status of each star in our sample.

\subsection{Radial Velocity} \label{subsec:rv}

We ran the IRAF task {\it fxcor} which calculates $V_{\rm{RAD}}$ and {\it v} sin {\it i} directly on the heliocentric-corrected, linear, and continuum subtracted spectra.

\begin{figure}
	\centering
	\includegraphics[width=0.5\textwidth]{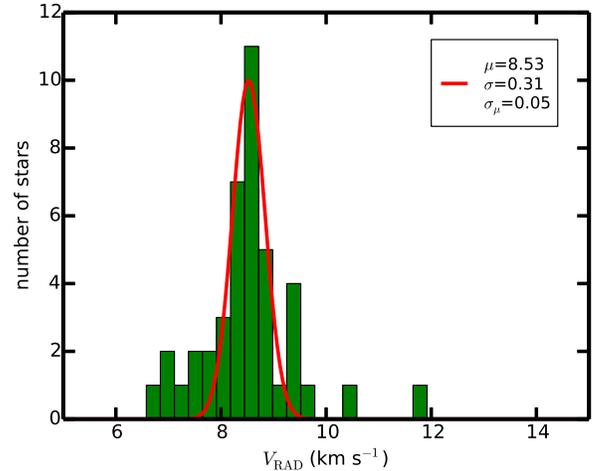}
	\caption{Radial velocity of m48f stars on night 1712n3. The mean and standard deviation of the Gaussian fit are 8.53 km s$^{-1}$ and 0.31 km s$^{-1}$, respectively.}
	\label{fig:rv_m48f1}
\end{figure}

We took spectra of $V_{\rm{RAD}}$ standards on each night to perform an external check on the wavelength calibration, with the exception of 1712n5 due to bad weather, as noted in Table \ref{tab:obs}. Reassuringly, with only a few exceptions like 1712n1, the large majority of measured $V_{\rm{RAD}}$ are within 2$\sigma$ of the literature values for the vast majority of nights. For M48 stars observed on more than one night, we measured $V_{\rm{RAD}}$ independently on each night. As an example, Figure \ref{fig:rv_m48f1} shows the $V_{\rm{RAD}}$ distribution of the m48f stars from night 1712n3. Typical errors for individual stars are 0.5--0.9 km s$^{-1}$. A Gaussian fit (dashed line) to the data yields a mean $V_{\rm{RAD}}$ of 8.53 $\pm$ 0.05 km s$^{-1}$ ($\sigma_{\mu}$, and $\sigma$=0.31 km s$^{-1}$). Table \ref{tab:obs} shows the results from similar fits to all configurations on all nights.

\subsection{Binarity}

Binarity can lead to misleading measurement of rotational velocity ({\it v} sin {\it i}) and equivalent width and, thus, abundance. For example, an indeterminate amount of contaminating flux from a secondary may lead to an indeterminate underestimation of equivalent width. So we have attempted to identify binaries and then eliminate them from subsequent analysis, where appropriate to do so. As discussed in Section \ref{sec:obs}, stars with $\bv\ >$ 0.40 mag were selected initially for spectroscopic follow-up if the $UBVRI$ photometry placed them on the apparent cluster single-star fiducial sequence. We have applied two additional criteria to help us determine binarity.

First, we compared $V_{\rm{RAD}}$ of the same configuration from different nights. All configurations were observed on at least two nights, and a few were observed on three nights, except m48rg, which was observed just once (see Table \ref{tab:obs}). If both (or all three) $V_{\rm{RAD}}$ measures for a given star agree to within 2$\sigma$, defined using the largest $\sigma$, we marked the star as a single star; if at least one measure disagreed by more than 2$\sigma$, we marked it as a binary; if there were ambiguities, we left a question mark. The second criterion evaluates the power spectrum from {\it fxcor}. Spectroscopic binaries have two peaks (or more) in the power spectrum. We did this separately for each night, so each star has binarity information for at least two nights. This also precludes confusion due to co-addition of binary spectra after orbital motion has shifted the spectra.

Under most circumstances, the second criterion agreed with the first. However, if the secondary is much fainter than the primary, the power spectrum might not be able to see enough flux to create a second peak. So, if the $V_{\rm{RAD}}$ are robustly different, i.e. the individual $V_{\rm{RAD}}$ errors were small compared to the differences in $V_{\rm{RAD}}$, we labeled the star as a binary.

\subsection{Membership and Final Cluster Radial Velocity } \label{subsec:mem_radv}

To identify stars consistent with single-star membership, we compared the $V_{\rm{RAD}}$ of individual stars to the average $V_{\rm{RAD}}$ of M48 as follows. For each configuration for each night, we chose a subsample that satisfied the following criteria: 1) single star according to the above combined binarity criteria, 2) {\it v} sin {\it i} $<$ 20 km s$^{-1}$, and 3) $\sigma_{V_{\rm{RAD}}} <$ 1.0 km s$^{-1}$. To this subsample, we then fit a Gaussian to the $V_{\rm{RAD}}$ distribution and calculated the average $V_{\rm{RAD}}$ and standard deviation. We initially ignored nights where our standard measurements did not agree well with literature and the m48vb1 and m48vb2 configurations because the luminous stars at the main-sequence turnoff have very high {\it v} sin {\it i}, leading to large uncertainties in the stellar $V_{\rm{RAD}}$. The weighted mean $V_{\rm{RAD}}$ from all of the considered configurations from the nights is 8.399 $\pm$ 0.037 km s$^{-1}$ ($\sigma_{\mu}$, and $\sigma$ = 0.099 km s$^{-1}$). We adopted this value temporarily as the average $V_{\rm{RAD}}$ for M48 ($<V_{\rm{RAD}}>$). We then shifted the average $V_{\rm{RAD}}$ from each and every configuration from the full sample to match this initial cluster $<V_{\rm{RAD}}>$, and combined the spectra from separate nights to get higher signal-to-noise ratio (S/N) spectra for each configuration. We then ran the {\it fxcor} task once again on the combined spectra. As in section \ref{subsec:rv}, we then fit a Gaussian in the $V_{\rm{RAD}}$ distribution for each configuration but only to the single stars. Finally, treating each configuration separately, we marked the stars within 2$\sigma$ of the mean as members, those between 2 and 3 $\sigma$ as uncertain (``?"), and those outside 3$\sigma$ as not-single members. They could be nonmembers, or member binaries whose binarity was not detected by the above techniques (note that these stars all lie less than 0.75 mag brighter than the left-edge fiducial).

\begin{figure}
	\centering
	\includegraphics[width=0.5\textwidth]{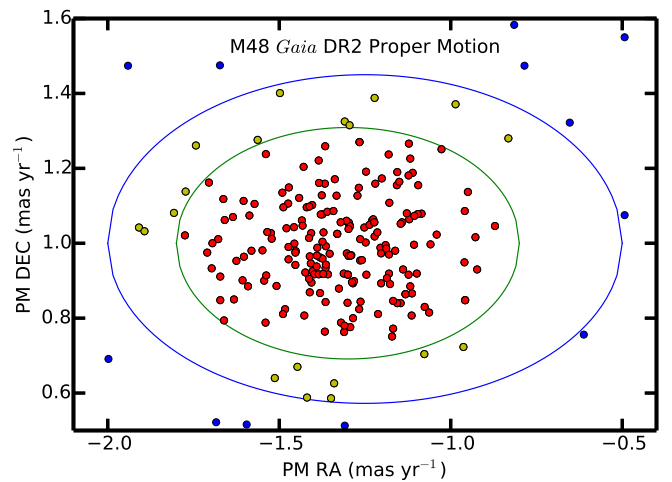}
	\caption{M48 proper motion (PM) membership determination using PM in both R.A. and decl. directions (mas yr$^{-1}$)), for our full Hydra sample. We define the center red circles as M48 PM members, the yellow circles as stars of uncertain PM membership, and the blue circles as PM nonmembers.}
	\label{fig:m48mem_gaia}
\end{figure}

\begin{figure}
	\centering
	\includegraphics[width=0.5\textwidth]{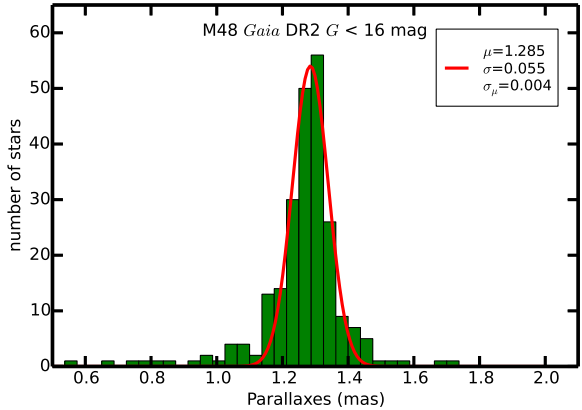}
	\caption{The parallax (mas) distribution for our full Hydra sample. We define stars that fall within 2$\sigma$ of the Gaussian fit as parallax members, those between 2$\sigma$ and 3$\sigma$ as having undetermined membership from parallax, and those beyond 3$\sigma$ as nonmembers. The mean and standard deviation of the parallax for M48 Hydra stars are 1.285 mas and 0.055 mas from the Gaussian fit.}
	\label{fig:m48plx_gaia}
\end{figure}

\begin{figure}
	\centering
	\includegraphics[width=0.5\textwidth]{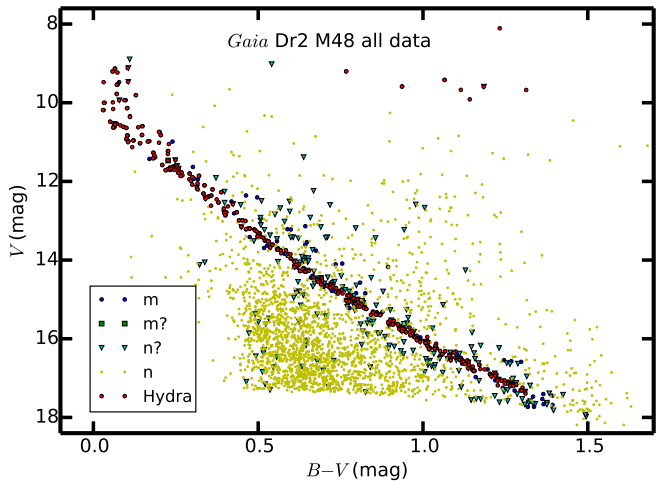}
	\caption{Color-magnitude diagram of M48 stars with Gaia membership. The blue circles are Gaia members, the green squares are uncertain members, the cyan triangles are uncertain nonmembers, and the yellow dots are nonmembers. The red circles are our observations of M48 stars using Hydra.}
	\label{fig:gaia_hydra_CMD}
\end{figure}

For further evidence of membership, we also considered the Gaia data, using proper motion ($P\mu_{\rm{R.A.}}$ \& $P\mu_{\rm{decl.}}$) and parallax from the Gaia DR2\citet[Gaia Collaboration][]{Gaia16} full degree that covers the Hydra field with a $G$ magnitude cut at $G$ = 17.2 mag, slightly fainter than the faint limit of our Hydra sample, and matched with our $UBVRI$ photometry. Figure \ref{fig:m48mem_gaia} shows Hydra M48 proper motion members selected using $P\mu_{\rm{R.A.}}$ and $P\mu_{\rm{decl.}}$ criteria. Parallax ($\pi$) was considered independently of proper motion. Figure \ref{fig:m48plx_gaia} shows a histogram of the number of stars versus $\pi$, where the cluster members clearly stand out from the other stars. Based on a Gaussian fit, we marked stars within 2$\sigma$ as members, between $2\sigma$ and $3\sigma$ as stars with uncertain membership, and stars outside 3$\sigma$ as nonmembers. We also considered two stars falling outside the $3\sigma$ region that had unusually large astrometric errors; neither has convincing evidence of membership, either from Gaia data or our $V_{\rm{RAD}}$ data, and are designated ``sn'' below. Given the frequency of stars outside the $\pi$ interval 1.05--1.50 mas, we estimate that the group identified as members may contain of the order of three nonmembers.

\begin{figure}
	\centering
	\includegraphics[width=0.5\textwidth]{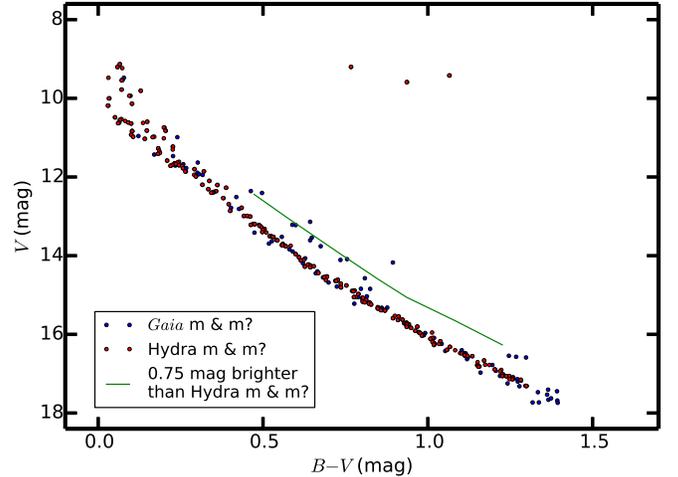}
	\caption{A comparison of M48 Gaia members and likely members (blue circles) to Hydra members and likely members (red circles). The green line is 0.75 mag brighter than Hydra fiducial.}
	\label{fig:gaia_hydra_m_CMD}
\end{figure}

How well did our photometric selection procedure pick out members? Figure \ref{fig:gaia_hydra_CMD} shows the stars observed with Hydra (red dots) with membership status using only Gaia membership information (no $V_{\rm{RAD}}$ information; blue dots are Gaia members, ``m"; green squares leaning toward membership, ``m?"; inverted triangles leaning toward nonmembership, ``n?"; and yellow dots nonmembers, ``n"). Reassuringly, the few instances of m? are near the fiducial sequence, while the vast majority of n? and n are scattered away from it. Our photometric method eliminated a good number of n that lie on the fiducial, but it also threw out of the order of 50 m on or very near the fiducial. This compares favorably to the number of stars observed with Hydra (192) whose final designation (below) is m (163) or m? (29). Although the photometric method also (deliberately) ignored potential high-q binary members, the number of such stars that were not observed with Hydra (15) is {\it vastly} outnumbered by the number of n? and n that lie up to 0.75 mag brighter than the fiducial. We can see the lack of a significant high-q binary sequence in Figure \ref{fig:gaia_hydra_m_CMD}.

\begin{figure*}
	\centering
	\includegraphics[width=0.7\textwidth]{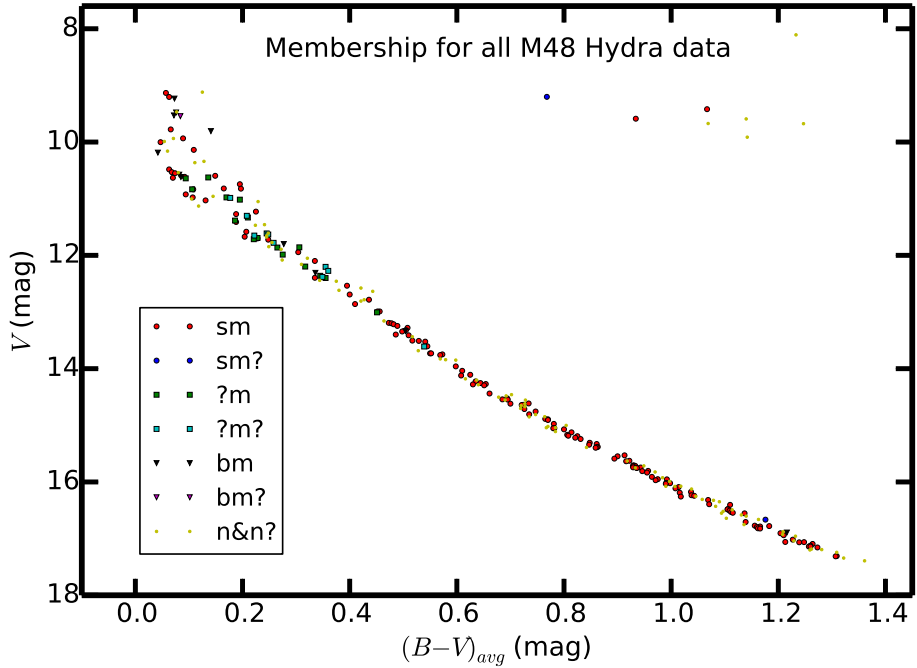}
	\caption{Color-magnitude diagram of M48 stars with membership and multiplicity information. The red circles are single members, the blue circles are likely single members, the green squares are members with unknown binarity, the light blue squares are likely members with unknown binarity, the black triangles are binary members, the magenta triangles are binaries of uncertain membership, and the small yellow dots are nonmembers and likely nonmembers.}
	\label{fig:cmd_BV}
\end{figure*}

We combined all of the $V_{\rm{RAD}}$, $P\mu_{\rm{R.A.}}$, $P\mu_{\rm{decl.}}$, and $\pi$ information to make a final decision on placing each star into one of the following categories: single-star member (sm), single-star nonmember (sn), binary member (bm), binary nonmember (bn), uncertain multiplicity (?m, ?n), and uncertain membership (``likely" member: sm?, ?m?, bm?; ``likely" nonmember: n? etc.). This results in 152 sm, 11 bm, 16 ?m, 56 sn, 28 sn?, 2 sm?, 1 bm?, 5 bn, 10 ?m?, 3 ?n, and 3 ?n?. Figure \ref{fig:cmd_BV} ($V$ versus $\bv$ CMD) and Table \ref{tab:atmosphere} show the final M48 membership and multiplicity determinations.

In more detail, our final determination of membership was carried out as follows. As discussed above, for each star, we assigned a membership status of ``y,'' ``n,'' or ``?'' to each of the following four criteria: photometry, $V_{\rm{RAD}}$, $P\mu$, and parallax.  (Recall that all stars have status ``y'' for photometry, since they were selected this way to begin with.)  Then, membership status was treated a bit differently for each of the three binarity cases (s,b,?).  For single stars, status ``sm'' was assigned if all four criteria had a ``y'' (130 stars) or if three criteria had a ``y'' and one had a ``?'' (22 stars).  Status ``sm?'' was assigned if two criteria had a ``?'' (two stars).  Status ``sn?'' was assigned if one criterion had an ``n'' and the other three were ``y'' (21 stars) or if one criterion had an ``n,'' one had a ``?,'' and the other two had a ``y'' (seven stars).  Finally, status ``sn'' was assigned if at least two criteria had an ``n'' (56 stars).  For binary (or multiple) stars, the radial velocity criterion was ignored.  Status ``bm'' was assigned if all three (remaining) criteria had a ``y'' (11 stars), ``bm?'' if one criterion had a ``?'' and two had a ``y'' (one star), and ``bn'' if at least one criterion had an ``n'' (five stars).  Binarity status ``?'' was treated as an intermediate case, and radial velocities were again included for consideration.  Status ``?m'' was assigned if all four criteria had a ``y'' (15 stars) or if three had a ``y'' and the fourth had a ``?'' in a category other than radial velocity (one star).  Status ``?m?'' was assigned if the radial velocity criterion had an ``n'' and the other three had ``y'' (eight stars), or if for the other three criteria, one had a ``?'' and two had a ``y'' (one star); or if the radial velocity criterion had a ``?'' and the other three had a ``y'' (one star).  Status ``?n?'' was assigned if two criteria had a ``?'' and two had a ``y'' (one star) or if one criterion had an ``n'', one had a ``?,'' and two had a ``y'' (three stars).  Finally, status ``n'' was assigned if at least two criteria had an ``n'' (two stars).

\begin{figure}
	\centering
	\includegraphics[width=0.5\textwidth]{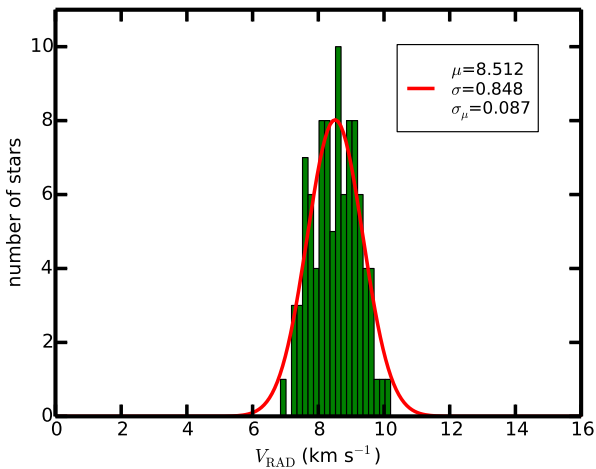}
	\caption{Final Radial Velocity of M48. The mean and standard deviation of the Gaussian fit are 8.512 km s$^{-1}$ and 0.848 km s$^{-1}$, respectively.}.
	\label{fig:Vrad_final}
\end{figure}

To determine a final $V_{\rm{RAD}}$ and {\it v} sin {\it i} for each star and to determine the final cluster average $V_{\rm{RAD}}$, we applied the procedure described at the beginning of section \ref{subsec:mem_radv} once again and used the same criteria: a) must be sm based upon our final determination, Gaia data included, b) {\it v} sin {\it i} $<$ 20 km s$^{-1}$, and c) $\sigma_{V_{\rm{RAD}}}\ <$ 1.0 km s$^{-1}$. Again, a Gaussian was fit to each configuration from each night. The weighted mean $V_{\rm{RAD}}$ is 8.376 $\pm$ 0.061 km s$^{-1}$ ($\sigma_{\mu}$, and $\sigma$ = 0.137 km s$^{-1}$). After shifting each configuration onto 8.376 km s$^{-1}$ and combining the spectra from separate nights, we ran {\it fxcor} on the combined spectra and reevaluated the $V_{\rm{RAD}}$ and {\it v} sin {\it i} of each star. We fit a Gaussian distribution function to all of the qualifying stars to arrive at a {\it final cluster average $V_{\rm{RAD}}$ of 8.512 $\pm$ 0.087 km s$^{-1}$} ($\sigma_{\mu}$, and $\sigma$ = 0.848 km s$^{-1}$; shown in Figure \ref{fig:Vrad_final}).

\begin{longrotatetable}
	\begin{deluxetable*}{cccccccccccccccccccccc}
		\label{tab:atmosphere}
		\tablecaption{Parameters and Metallicity for M48 stars}
		\tabletypesize{\tiny}
		\setlength{\tabcolsep}{3pt}
		\renewcommand{\arraystretch}{1.0}
		\tablehead{
			\colhead{Star Id} & \colhead{R.A.} & \colhead{Decl.} &
			\colhead{$V^1$} & \colhead{$\bv^1$} &
			\colhead{$V_{\rm{RAD}}^2$} & \colhead{$\sigma^2$} &
			\colhead{$V_{\rm{RAD}}^3$} & \colhead{$\sigma^3$} &
			\colhead{{\it v} sin {\it i}$^4$} & \colhead{$\sigma^4$} & \colhead{$H_{\alpha}$?$^5$} &
			\colhead{$(\bv)_{\rm{eff}}^6$} & \colhead{$\sigma^6$} & 
			\colhead{$T_{\rm{eff}}^7$} & \colhead{$\sigma^7$} & 
			\colhead{log {\it g}$^7$} & \colhead{$V_{\rm t}^7$}&
			\colhead{[Fe/H]$^8$} & \colhead{$\sigma_{\mu}^8$} &
			\colhead{S/N$^9$} & \colhead{mem$^10$}\\
			&h m s&$\arcdeg\ \arcmin\ \arcsec$& mag & mag & \colhead{km s$^{-1}$} & \colhead{km s$^{-1}$} & \colhead{km s$^{-1}$} & \colhead{km s$^{-1}$} &
			\colhead{km s$^{-1}$} & \colhead{km s$^{-1}$} & & mag & mag &
			\colhead{K} & \colhead{K} & 
			\colhead{} & \colhead{km s$^{-1}$}&
			\colhead{dex} & \colhead{dex} &
			\colhead{} & \colhead{}
		} 
		\startdata
		\hline
		2001&8 12 08.38&-6 05 38.7&8.108&1.233&0.84&1.35&--&--&16&0.7&no&--&--&--&--&--&--&--&--&185&sn \\
		2002&8 13 44.84&-5 48 00.6&9.202&0.767&3.54, 4.38&5.42, 1.67&1.35&1.78&14&0.4&no&0.768&0.002&5436&6.2&4.55&0.80&--&--&1063&sm? \\
		2003&8 14 28.12&-5 42 16.1&9.420&1.065&11.49, 11.71&1.88, 1.34&9.04&1.43&16&0.6&no&1.067&0.002&4549&4.8&4.69&0.80&--&--&959&sm \\
		2004&8 13 35.43&-5 53 02.1&9.588&0.936&10.4, 11.67&1.21, 2.58&8.70&1.93&21&0.6&no&0.934&0.004&4912&10&4.64&0.80&--&--&904&sm \\
		2005&8 13 38.04&-6 01 32.2&9.592&1.184&31.40, 30.27&2.36, 1.75&27.82&1.70&21&0.6&no&1.140&0.049&--&--&--&--&--&--&939&sn \\
		2006&8 12 51.29&-5 50 50.8&9.673&1.115&92.57, 95.53&2.80, 1.54&92.35&1.49&18&0.6&no&1.069&0.029&--&--&--&--&--&--&906&sn \\
		2007&8 12 36.49&-5 39 50.3&9.676&1.313&33.40, 37.15&3.10, 1.67&33.81&1.84&22&0.6&no&1.247&0.067&--&--&--&--&--&--&948&sn \\
		2008&8 14 26.34&-5 44 34.5&9.914&1.142&32.73, 28.94&1.88, 2.48&27.17&2.40&29&1.0&no&--&--&--&--&--&--&--&--&867&sn \\
		2009&8 14 15.50&-5 43 15.8&9.119&0.104&8.81, 15.44&5.56, 5.60&10.63&3.31&--&--&--&0.125&0.020&--&--&--&--&--&--&914&sn? \\
		2010&8 13 44.24&-5 48 48.6&9.133&0.065&15.21, 9.77&1.38, 3.51&8.67&1.31&34&4.2&no&0.057&0.010&--&--&--&--&--&--&518&sm \\
		2011&8 13 08.55&-5 38 35.6&9.204&0.058&14.74, 13.11&4.43, 9.03&9.63&6.61&300&--&yes&0.063&0.012&--&--&--&--&--&--&807&sm \\
		2012&8 13 46.65&-5 44 52.3&9.233&0.073&1.16, 14.27&5.67, 3.24&--&--&--&--&--&--&--&--&--&--&--&--&--&740&bm \\
		2013&8 13 52.98&-5 42 46.5&9.468&0.105&16.38, 15.95&3.69, 9.49&11.33&7.39&--&--&--&0.077&0.020&--&--&--&--&--&--&704&sn? \\
		2014&8 13 05.38&-5 45 00.5&9.478&0.031&17.22, -11.53&3.86, 6.79&--&--&--&--&--&0.076&0.009&--&--&--&--&--&--&592&bm \\
		2015&8 13 28.67&-5 48 15.0&9.530&0.073&-29.4,-17.87(45.16)$^{*}$&3.96,4.24(6.90)$^{*}$&--&--&29(31)$^{*}$&3.2(5.1)$^{*}$&no&0.072&0.009&--&--&--&--&--&--&661&bm \\
		2016&8 12 58.57&-5 34 08.1&9.543&0.071&38.88, 19.34&2.32, 3.08&--&--&20&4.8&no&0.084&0.009&--&--&--&--&--&--&719&bm? \\
		2017&8 13 23.22&-5 45 23.0&9.777&0.071&11.69, 13.58&0.86, 4.05&10.63&3.29&23&2.5&no&0.066&0.007&--&--&--&--&--&--&579&sm \\
		2018&8 13 39.66&-5 47 14.6&9.807&0.129&20.33, 4.90&4.79, 3.52&--&--&--&--&--&0.141&0.029&--&--&--&--&--&--&664&bm \\
		2019&8 13 04.96&-5 53 04.8&9.935&0.094&20.67, 14.96&9.76, 7.88&11.48&8.09&250&--&yes&0.089&0.006&--&--&--&--&--&--&561&sm \\
		2020&8 13 49.00&-5 44 23.7&9.937&0.078&-12.97, -1.19&4.35, 6.92&-7.20&5.93&--&--&--&0.071&0.019&--&--&--&--&--&--&571&?n \\
		2021&8 13 26.60&-5 49 53.8&9.990&0.059&26.75, 6.46&5.49, 4.98&--&--&--&--&--&0.054&0.008&--&--&--&--&--&--&554&bn \\
		2022&8 13 43.39&-5 41 33.7&10.002&0.032&18.85, 15.86&6.26, 9.34&12.95&9.06&230&--&yes&0.047&0.006&--&--&--&--&--&--&515&sm \\
		2023&8 13 40.40&-5 42 20.1&10.138&0.102&10.93, 7.13&4.81, 4.69&6.86&2.95&150&--&yes&0.109&0.013&--&--&--&--&--&--&570&sm \\
		2024&8 13 54.40&-5 58 47.6&10.160&0.060&11.47, 12.48&3.12, 9.77&8.60&5.86&--&--&--&0.060&0.003&--&--&--&--&--&--&531&sn \\
		2025&8 14 03.19&-5 41 44.5&10.187&0.029&20.83, 1.83&3.02, 2.74&--&--&40&6.0&no&0.042&0.008&--&--&--&--&--&--&513&bm \\
		2027&8 13 12.17&-5 46 41.8&10.341&0.111&15.83, 10.39&4.55, 5.74&9.32&5.90&--&--&--&0.128&0.016&--&--&--&--&--&--&453&sn? \\
		2028&8 13 45.99&-5 46 01.9&10.364&0.111&2.25, 2.2&2.02, 2.49&-0.48&2.18&46&5.2&no&--&--&--&--&--&--&--&--&492&sn \\
		2029&8 12 08.26&-6 03 16.0&10.483&0.051&17.78, 11.65&4.55, 6.35&11.18&6.60&230&--&yes&0.063&0.008&--&--&--&--&--&--&376&sm \\
		2030&8 14 20.28&-5 39 57.4&10.531&0.071&3.64, 1.93&4.24, 3.90&-0.39&3.12&42&6.5&no&0.068&0.008&--&--&--&--&--&--&400&sm \\
		2031&8 13 17.60&-5 41 13.4&10.540&0.064&23.3, 22.13&6.94, 9.67&19.80&9.21&--&--&--&0.080&0.013&--&--&--&--&--&--&458&sn? \\
		2032&8 14 02.41&-5 56 46.8&10.550&0.066&4.61, 1.09&2.00, 3.12&-0.28&2.84&20&1.9&no&0.074&0.006&--&--&--&--&--&--&376&sm \\
		2033&8 13 43.25&-5 45 53.1&10.576&0.082&37.77, -19.73&2.46, 4.49&--&--&52&8.3&no&--&--&--&--&--&--&--&--&433&bm \\
		2035&8 13 52.12&-5 54 20.3&10.596&0.148&20.95, 16.49&5.55, 9.10&14.44&7.34&280&--&yes&0.149&0.014&--&--&--&--&--&--&404&sm \\
		2036&8 13 13.82&-5 56 38.2&10.614&0.091&21.40, 14.18&9.01, 5.75&14.56&8.34&230&--&yes&0.092&0.012&--&--&--&--&--&--&447&sm \\
		2037&8 13 19.73&-5 33 37.2&10.617&0.063&-0.07, 30.72&2.35, 2.33&--&--&24&2.1&no&0.085&0.015&--&--&--&--&--&--&437&bm \\
		2038&8 13 09.50&-5 27 01.1&10.625&9.999&18.78, 9.43&7.59, 5.49&8.86&6.05&--&--&--&0.136&0.013&--&--&--&--&--&--&437&?m \\
		... & ... & ... & ... & ... & ... & ... & ... & ... & ... & ... & ... & ... & ... & ... & ... & ... & ... & ... &... &...&... \\
		\hline
		\enddata
		\tablecomments{1. $V$ magnitude and $\bv$ color from our M48 photometry.
			2. Radial velocity ($V_{\rm{RAD}}$) and errors in km s$^{-1}$ reported for individual nights.
			3. $V_{\rm{RAD}}$ and errors in km s$^{-1}$ measured by the combined spectra for single stars and stars with uncertain multiplicity.
			4. Rotational velocity ({\it v} sin {\it i}) and errors in km s$^{-1}$. 
			5. Whether {\it v} sin {\it i} are measured using $H_{\alpha}$ or by averaging lines between 6600 \AA and 6800 \AA.
			6. Averaged ($\bv$) and standard deviation by using all 10 possible color combinations from $UBVRI$.
			7. Stellar atmosphere parameters derived from section \ref{subsec:atmosphere}: $T_{\rm{eff}}$ \& $\sigma_{T_{\rm{eff}}}$ in K, log {\it g}, and $V_{\rm t}$ in km s$^{-1}$.
			8. [Fe/H] and $\sigma_{\mu}$([Fe/H]) for individual stars based on calculations in section \ref{subsec:feh}, for single-member stars that satisfy the {\it v} sin {\it i} and $\sigma_{T_{\rm{eff}}}$ criteria only.
			9. Signal-to-noise ratio of the combined spectra of all nights of the star.
			10. Binarity \& membership determination from section \ref{sec:rv}. sm: single member; bm: binary member; ?m: member of uncertain multiplicity; sn: single star nonmember; sn?: single star likely non-member; sm?: single star likely member; bm?: binary likely member; bn: binary nonmember; ?m?: likely member of uncertain multiplicity; ?n: nonmember of uncertain multiplicity; ?n?: likely nonmember of uncertain multiplicity.
			$^{*}$ $V_{\rm{RAD}}$ and {\it v} sin {\it i} for secondary star measured from \it fxcor.}
	\end{deluxetable*}
\end{longrotatetable}

\begin{figure*}
	\centering
	\includegraphics[width=0.7\textwidth]{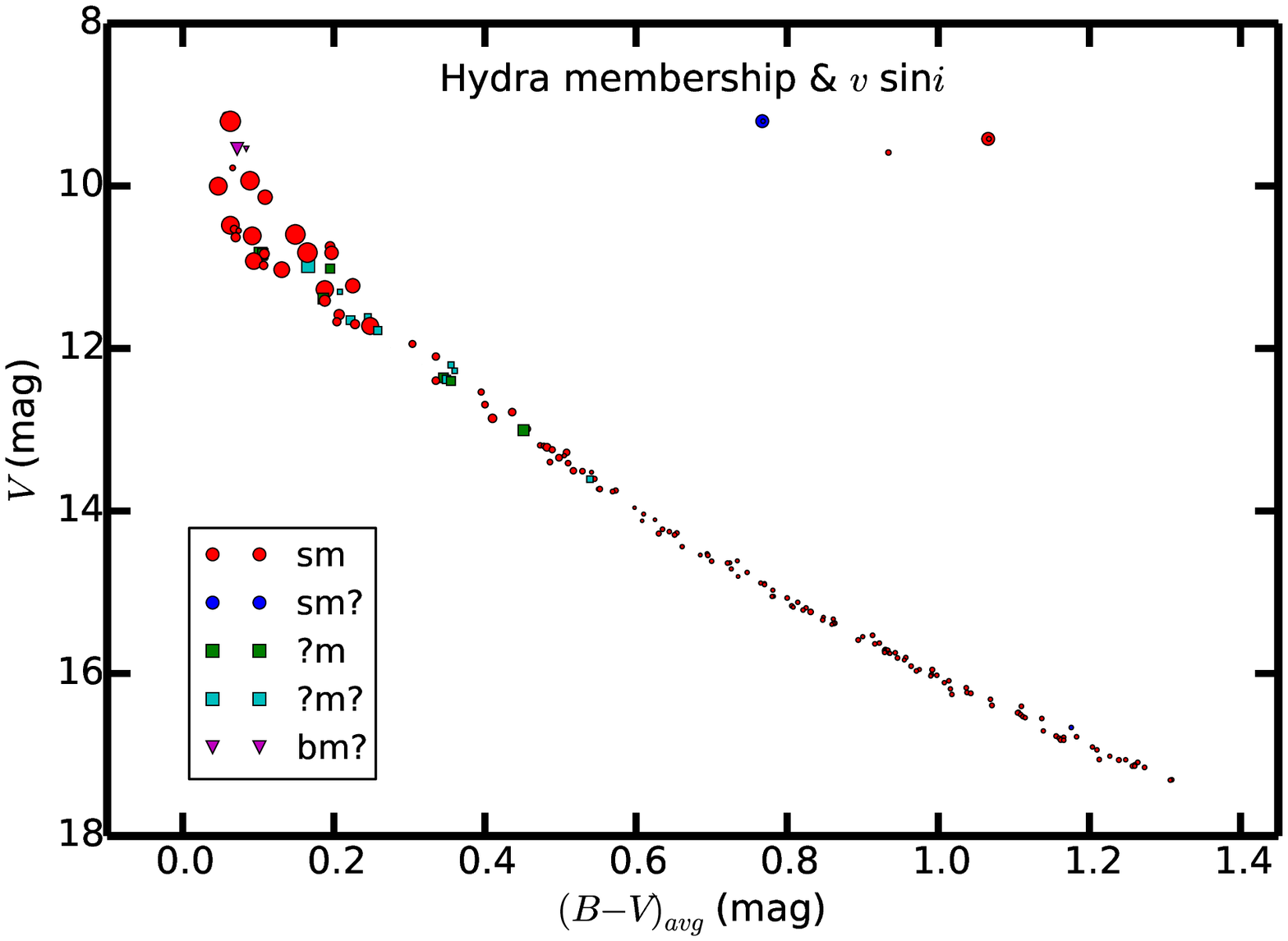}
	\caption{Color-magnitude diagram of M48 members and likely members with rotational velocity. The marker size is proportional to $\sqrt{v\ \rm{sin}\ {\it i}}$.}
	\label{fig:cmd_rot}
\end{figure*}

To determine final $V_{\rm{RAD}}$ and {\it v} sin {\it i} for each star, we shifted $V_{\rm{RAD}}$ from each and every configuration onto the new cluster average of 8.512 km s$^{-1}$ (these final $V_{\rm{RAD}}$ from each individual night are reported in Table \ref{tab:atmosphere}), combined spectra from the same configuration, and determined the final $V_{\rm{RAD}}$ and {\it v} sin {\it i} for each star (reported in Table \ref{tab:atmosphere}). The reader should be cautious about the $V_{\rm{RAD}}$ of those stars that have multiplicity status ``?". Similarly, {\it v} sin {\it i} may not be accurate for stars that are not single. Furthermore, even though we report the {\it v} sin {\it i} produced by {\it fxcor}, values far below our resolution limit of roughly 10--12 km s$^{-1}$ are uncertain, even for single stars. A conservative interpretation might treat values less than 10--12 km s$^{-1}$ as upper limits of 12 km s$^{-1}$. Star 2015 has two clear peaks from {\it fxcor} and we are able to measure the $V_{\rm{RAD}}$ and {\it v} sin {\it i} for each separate peak, so we report $V_{\rm{RAD}}$ and {\it v} sin {\it i} for both stars in Table \ref{tab:atmosphere} (secondary shown in parentheses). For some very hot and/or rapidly rotating stars, {\it fxcor} was unable to determine {\it v} sin {\it i}. In these cases, we obtained a rough estimate of {\it v} sin {\it i} (to within roughly 10--20 $\%$) by synthesizing the $H_{\alpha}$ line from 6515 to 6610 \AA\ (indicated as ``yes'' in column $H_{\alpha}$ of Table \ref{tab:atmosphere}) using MOOG \citet[Sneden et al.][]{Sneden73}. In choosing the best-fitting value of {\it v} sin {\it i}, we were guided by similar syntheses of the most rapidly rotating stars of similar spectral type that had {\it fxcor}-determined values of {\it v} sin {\it i}.  We do not report {\it v} sin {\it i} from $H_{\alpha}$ for stars of uncertain binarity or membership. Figure \ref{fig:cmd_rot} presents the {\it v} sin {\it i} of M48 members and likely members (m, m?) in the $V$ versus $\bv$ CMD, with the symbol size proportional to $\sqrt{v\ \rm{sin}\ {\it i}}$, where {\it v} sin {\it i} ranges from 6 to 300 km s$^{-1}$. Note that for configurations observed on more than one night, Table \ref{tab:atmosphere} shows final $V_{\rm{RAD}}$s and errors from the combined spectra of multiple nights only for single stars and stars with uncertain multiplicity. The shifts of final $V_{\rm{RAD}}$ from intermediate $V_{\rm{RAD}}$ are very small, always less than 0.2 km s$^{-1}$ for all single stars. 

We compare our cluster radial velocity to three previous reports of $V_{\rm{RAD}}$ in M48. We are in good agreement with Wallerstein et al.\citet{Wallerstein63}, who report 8.9 km s$^{-1}$ (no error reported) based on three giants, and who suspect their $V_{\rm{RAD}}$ are systematically too high by up to 1 - 2 km s$^{-1}$. Our value is just slightly higher than that of Mermilliod et al.\citet{Mermilliod08a}, who report 7.70 $\pm$ 0.18 km s$^{-1}$ (``error") from four giants of which two are SB. Note that of the eight giants observed by us, two are sm, one is sm?, and five are sn. Geyer \& Nelles\citet{Geyer85} report 5.7 $\pm$ 1.3 km s$^{-1}$ (m.e.) from 21 stars, which have a range in $V_{\rm{RAD}}$ of -20 to 42 km s$^{-1}$ with errors ranging from 2.6 to 6.8 km s$^{-1}$ (two stars have larger errors). Their $V_{\rm{RAD}}$ distribution peaks at 6--10 km s$^{-1}$, in agreement with our result.

Finally, we comment on the possible relation between the broadening of the turnoff and stellar rotation. Evidence that cluster turnoffs in the CMD can be much wider than the single-star fiducial at lower mass has been around for a very long time; compare, for example, the very thin single star fiducial in Praesepe to the much wider cluster turnoff\citet[Johnson][]{Johnson52}. Such ``extended'' (broadened) main-sequence turnoffs (eMSTOs) have been observed in many more clusters, such as in most massive clusters in the Large and Small Magellanic Clouds (MC) with age $<$ 2.5 Gyr (e.g. Mackey \& Broby Nielsen\citet{Mackey07}; Mackey et al.\citet{Mackey08}; Correnti et al.\citet{Correnti17}), some of which (age $<$ 700 Myr) show ``split'' (bimodal) main-sequences (Milone et al.\citet{Milone13}; Goudfrooij et al.\citet{Goudfrooij14}; Li et al.\citet{Li17}). Aided by Gaia DR2 membership information and photometry, increasingly, Milky Way (MW) open clusters are also found to exhibit eMSTOs\citet[Cordoni et al.][]{Cordoni18}, even though they generally are much less massive than the MC clusters. Deciphering the origin of eMSTOs is thus becoming of increasing interest.

One contributor could be binarity, but binaries should broaden the entire main-sequence, not just the turnoff, so unless the binary fraction is much larger for more massive stars (see also Section \ref{subsec:binary}), other contributors may be important. In fact, M48 exhibits a modest eMSTO even among stars identified as single (red disks in Figure \ref{fig:cmd_rot}). Another posited contributor among clusters with ages 1--3 Gyr has been variability\citet[Salinas et al.][]{Salinas16}; however, the number of variable stars in the MC cluster NGC 1846 may not be sufficiently large\citet[Salinas et al.][]{Salinas18}. Another possibility is age spreads due to prolonged star formation or multiple epochs of star formation, in which case, the MC clusters could be younger analogs of MW globular clusters that show multiple populations\citet[Keller et al.][]{Keller11}. However, it is expected that only very massive star clusters can create multiple populations\citet[e.g. D'Ercole et al.][]{DErcole08}, which would exclude almost all MW open clusters, and evidence such as the existence of multiple main-sequences that attest to the multiple populations in MW globular clusters has yet to be discovered in open clusters. Furthermore, the implied age spreads can be absurdly large; for example, up to 500 Myr in an open cluster (NGC 5822) with an age of 900 Myr\citet[Figure 3 of Sun et al.][]{Sun19}. By contrast, evidence suggests an absence of primordial cluster gas after 4 Myr\citet[Hollyhead et al.][]{Hollyhead15} and no star formation in clusters older than 10 Myr (Elmegreen \& Efremov\citet{Elmegreen97}; Niederhofer et al.\citet{Niederhofer16}). Other explanations include metallicity variations\citet[Milone et al.][]{Milone15} and braking of rapid rotators\citet[D'Antona et al.][]{DOAntona17}. Perhaps the most promising explanation is a range in rotation rates among turnoff stars, which can broaden the MSTO through inclination angle and/or structural effects (von Zeipel\citet{von24}; Bastian \& de Mink\citet{Bastian09}).

Photometric\citet[Bastian et al.][]{Bastian17} and spectroscopic (Dupree et al.\citet{Dupree17}; Marino et al.\citet{Marino18}) studies provided some evidence that some eMSTOs in MC clusters have blue slow rotators and red fast rotators, consistent with early expectations\citet[Bastian \& de Mink][]{Bastian09}. However, more recent detailed models taking into account both inclination and structure effects predict little correlation between turnoff color and rotational {\it v} sin {\it i}\citet[Brandt \& Huang][BH15]{BH15}. For detailed discussion of the various issues and complexities we refer the reader to BH15, and comment here only on the interesting prediction listed above, and one more. In particular, BH15 find that in an observational color-magnitude diagram, the thickness (in color) of the eMSTO is small at younger ages ($<$ 500 Myr), then grows and peaks between 1 and 1.5 Gyr, and becomes thin at older ages. Consistent with both predictions, Figure \ref{fig:cmd_rot} shows a clear eMSTO of rather modest thickness and no discernible correlation between color and {\it v} sin {\it i}. We interpret the data of Sun et al.\citet{Sun19} for open cluster NGC 5882 in a similar way: at the turnoff ($G$ = 11--12 mag) the two most rapid rotators are redder and have {\it v} sin {\it i} = 230--250 km s$^{-1}$, but the next six most rapidly rotating stars are bluer and rotate only slightly less rapidly, with {\it v} sin {\it i} = 150--220 km s$^{-1}$.

\subsection{Cluster Binary Fraction} \label{subsec:binary}

The combination of our photometric data, our spectroscopic data, and the Gaia data enable us to examine the binary/multiple fraction of M48. None of the three m or m? giants showed evidence of binarity, but {\it fxcor} could have missed such evidence if the companion is much fainter, so we restrict attention to the main-sequence and turnoff stars. Since the analysis of rapid rotators is more challenging, we separate the sample into a ``hot'' subsample of stars rotating more rapidly that includes dwarfs with $\bv\ \leq $ 0.40 mag, and a ``cool'' subsample of stars rotating more slowly that includes dwarfs with $\bv\ >$ 0.40 mag (Figure \ref{fig:cmd_rot}). We do not consider stars with $\bv\ >$ 1.3 mag to match the cool limit of our Hydra sample. Here, we consider m and m? to be members and n and n? to be nonmembers.

We first consider the cool sample and begin with stars observed with Hydra. Recall that the fiducial was chosen photometrically as a thin, left-edge fiducial with deliberate intent to avoid binaries for Hydra observation. However, in principle, it is possible that some low-q binaries may reside on this photometric fiducial if the companion is too faint to contribute significantly to the total light detected. Nevertheless, of 123 members, 119 are single (96.7 \%), two are binary (1.6\%), and the other two have ? binarity/multiplicity status (1.6\%), illustrating that our technique was extremely successful in identifying single members.

To estimate the binary/multiple fraction, we must also consider those photometric binaries (defined as stars photometrically above the fiducial) that we specifically avoided for Hydra observation and any other members on the fiducial itself that were not observed with Hydra. For this purpose, we rely on membership as determined by the methods discussed above using the Gaia data alone. We find 63 members whose photometry places them on the fiducial and 19 stars that lie above the fiducial. Since we have no spectroscopic information on these 82 stars, we determine binarity/multiplicity using photometry as follows. For the 63 fiducial stars, if we assume the same fraction of binaries as in the Hydra fiducial sample, we conclude that 61 are single, one is b, and one is ?. We have no choice but to assume all 19 photometric binaries are binary/multiple. Note that, consistent with this assumption, nearly all of these putative binaries lie within 0.75 mag of the fiducial; the equal-mass binary sequence is shown in Figure \ref{fig:gaia_hydra_m_CMD} at 0.75 mag above the fiducial.  Roughly half of the 19 binary candidates lie close to this sequence, consistent with the idea that a range of q, not just q = 1, approach this ``equal-mass'' sequence.  One or possibly two stars lie above the sequence; these might be trinary/multiple members. We cannot rule out that other reasons may exist to displace a single star from the fiducial, but there are also no good reasons to believe that such displacements have occurred in our sample; we are bound by the absence of further information. Note also that for stars in this range in $\bv$, there is some evidence that rapid rotation is not a possible cause of such displacement: single rapid rotators in the Pleiades fall on the fiducial \citet[Soderblom et al.][]{Soderblom93a, Soderblom93b}. Note also the caveat that we might misidentify a binary as single if it has a sufficiently long period so that its radial velocity did not vary significantly between the dates of the two (or three) measurements and if the secondary flux is sufficiently low so as not to be detected by {\it fxcor}.

Summing up the numbers of the cool sample, of 205 members, 180 are single (87.8\%), 22 are binaries (10.7\%), and three are ? (1.5\%). Given our very limited number of epochs (two or three) it is impossible to attach errors to these numbers, as do a number of other studies that have multiple epochs of observations and determine binary orbits, and they are often able to make estimates of incompleteness (below).

Rapid rotation complicates analysis of the hot sample of more massive stars. In contrast to the cool sample, hot rapid rotators can be photometrically either on or off of the left-edge fiducial sequence (Figure \ref{fig:cmd_rot}). So we limit analysis to the 66 Hydra members and ignore the additional 16 Gaia members not observed with Hydra. Of the 66 Hydra members, 32 are single (48.5\%), 10 are binaries (15.2\%), and 24 are ? (36.4\%). If the ? are all binaries, then the hot binary fraction could be as high as 51.6\% and the combined hot+cold fraction would be 21\%. However, given the various uncertainties introduced by rapid rotation, it is difficult to ascertain whether these results truly differ from those of the cool sample. For example, it is also possible that all hot ? are single, in which case, the results of the cool and hot samples would be fairly similar. A mass-dependent binary fraction is sometimes seen; for example, Bohm-Vitense\citet{Bohm07} reports that the binary fraction in the Hyades increases from 26\% in K dwarfs to 87\% in A dwarfs. However, we can neither claim nor preclude a similar trend of increasing binary fraction with mass in M48.

The binary fractions in open clusters vary significantly. For the 100 Myr-old Pleiades, using 144 G and K dwarfs, Bouvier et al.\citet{Bouvier97} report a fraction of 28\% based on 22 binaries with separations 11--910 au and corrected for incompleteness, and Richichi et al.\citet{Richichi12} report 29\% from a smaller number of dwarfs of more varied spectral type. Bouvier et al.\citet{Bouvier97} point out the much higher binary fraction in some star forming regions such as Taurus-Auriga and Ophiuchus (Leinert et al.\citet{Leinert93}, Ghez et al.\citet{Ghez93}, Simon et al.\citet{Simon95}), and they suggest that cluster formation environment rather than cluster evolution is a more important factor: Pleiades is a dense cluster whereas the clouds are loose T Tauri associations. Mermilliod et al.\citet{Mermilliod08b} report a fraction of 20\% in FGK dwarfs of the 100 Myr-old Blanco 1, and Geller et al.\citet[WOCS study]{Geller10} report an incompleteness-corrected fraction of 24\% in the 150 Myr-old M35 for binaries with periods $<$ 10,000 days. Going to older ages than M48 (420 Myr), the 650 Myr-old Hyades has a much higher binary fraction (above). But Hole et al.\citet[WOCS study]{Hole09} report a fraction of 17\% for the 2.3 Gyr-old NGC 6819 ($P\ <$ 10,000 days, not corrected for incompleteness), and Geller \& Mathieu\citet[GM12, WOCS study]{Geller12} report an incompleteness-corrected fraction of 29\% for the 7 Gyr-old NGC 188 ($P\ <$ 10,000 days). Mathieu et al.\citet{Mathieu90} report a fraction of 9--15\% for the 4 Gyr-old M67 for $P\ <$ 1,000 days. For NGC 188, GM12 find 22\% for $P\ <$ 1,000 days. It is not at all clear that similar corrections apply to both clusters, but if they do, the corrected M67 fraction {\it might} be 12--20\%. In its younger days, NGC 188 may have had a smaller binary fraction, perhaps more comparable to that of M67, if evaporation favors single stars instead of binaries, as suggested by the models of Hurley et al.\citet{Hurley05}. On the other hand, destruction of binaries through internal cluster dynamics may also play an important role. Finally, Raghavan et al.\citet{Raghavan10} studied several hundred field stars and found a binary fraction of 19\% for $P\ <$ 10,000 days.

While a number of these clusters spanning 7 Gyr in age seem to show remarkably similar fractions, there also seem to be some exceptions: the Hyades (and the clouds) are much higher, and M48 might be marginally low. One possible distinguishing characteristic for the Hyades is its higher metallicity ([Fe/H] = +0.15 dex; Cummings et al.\citet{Cummings17}) compared to the other clusters ([Fe/H] = -0.2 to +0.05 dex; Hobbs et al.\citet{Hobbs90}; Friel \& Boesgaard\citet{Friel92}; Barrado y Navascues et al.\citet{Barrado01a}; Ford et al.\citet{Ford05}; Lee-Brown et al.\citet{Lee-Brown15}; Anthony-Twarog et al.\citet{Anthony18b}), but it is not clear whether this is related to binary fraction. For M48, given the various uncertainties it is not altogether clear that the binary fraction is actually low compared to the other clusters and the field. The cool sample (FGK dwarfs) does seem to have a low fraction (11\%) of binaries; although, inclusion of the hot sample (A dwarfs) might possibly bring the overall fraction up to 21\%. A possible distinguishing characteristic for M48 may be its richness: it is considerably less rich than M35, NGC 6819, M67, and NGC 188, but only slightly less rich than the Pleiades and Hyades, and M48 is richer than Blanco 1. Note that Blanco 1's binary fraction is also among the lowest listed above, so perhaps both M48 and Blanco 1 share richness (or absence thereof) as a common distinguishing feature.

\section{Metallicity} \label{sec:abun}

To determine stellar and cluster metallicities, we follow procedures very similar to those in Cummings et al.\citet{Cummings17}, which we briefly summarize here. These procedures include deriving precision cluster [Fe/H] based on as many isolated Fe I lines as possible in our spectral range, using as many carefully selected stars as possible covering as wide a range in $T_{\rm{eff}}$ as possible, and ensuring that we use only the range in $T_{\rm{eff}}$ for each line in which that line is well behaved. Cummings et al.\citet{Cummings17} were able to use Praesepe stars covering a range of 1700 K in $T_{\rm{eff}}$. For M48, we extend the range to 2500 K.

\subsection{Effective Temperature, Log g, and Microturbulence} \label{subsec:atmosphere}

We have adopted the following cluster input parameters, as derived from our $UBVRI$ photometric study (Paper II): distance $(m-M)_v = 9.47\ \pm\ 0.08$ mag, $age = 420\ \pm\ 30$ Myr, interstellar reddening $E(\bv)=0.05\ \pm\ 0.01$ mag, and metallicity [Fe/H] = -0.05 $\pm$ 0.03 dex.

We use $\bv$ colors to determine the effective temperature of our M48 dwarfs. To incorporate atmospheric information contained outside the $B$ and $V$ spectral ranges, and to reduce statistical and systematic errors, we use all 10 possible color combinations from $UBVRI$ to derive an effective, average $\bv$ for each star. For instance, we fit a polynomial to the $U$--$V$ vs. $\bv$ plot using the members in the M48 fiducial sequence, and then convert the $U$--$V$ for each star to the corresponding $\bv$ according to this relation. Similarly, we convert the $U$--$B$, $U$--$R$, $U$--$I$, $B$--$R$, $B$--$I$, $V$--$R$, $V$--$I$, and $R$--$I$ to $\bv$. Then we average the 10 $\bv$ colors to derive the final effective $\bv$ and $\sigma(\bv)$ (See Table \ref{tab:atmosphere}). Some stars lack measurements in certain bands, so for these stars we use only the measured colors to derive $\bv$.

To remain consistent with our previous studies (for example, Thorburn et al.\citet{Thorburn93}; Deliyannis et al.\citet{Deliyannis94},\citet{Deliyannis02},\citet{Deliyannis19}; Steinhauer \& Deliyannis\citet{Steinhauer04}; Anthony-Twarog et al.\citet{Anthony09},\citet{Anthony10},\citet{Anthony18a}; Maderak et al.\citet{Maderak13}), we have used the $(\bv)_0-[Fe/H]-T_{\rm{eff}}$ relation in equation (1) of Cummings et al.\citet{Cummings17}. For M48, we have assumed $E(\bv)=0.05$ mag and [Fe/H] = -0.05 dex, and, as usual, for the Hyades we assume [Fe/H]$_{\rm{Hyades}} = +0.15$ dex. $\sigma_{T_{\rm{eff}}}$ are calculated from $\sigma(\bv)$ based on error propagation. This relationship is valid for $T_{\rm{eff}}$ = [3500 K, 7750 K], which excludes many of our m48vb1 and m48vb2 stars that are hotter than 7500 K. However, none of these hotter stars meet the stringent selection criteria for metallicity determination defined in section \ref{subsec:feh}, so their exclusion does not affect the results of this study.

We determined the log {\it g} of each star from the Yonsei--Yale ($Y^2$; Demarque et al.\citet{Demarque04}) isochrones, adopting [Fe/H] = -0.05 dex, Z = 0.01618, Y = 0.26236, [$\alpha$/Fe] = 0.00, and an age of 420 Myr. Lastly, the microturbulence ($V_{\rm t}$) was calculated using the empirical relation for dwarfs of Edvardsson et al.\citet{Edvardsson93}, or 0.8 km s$^{-1}$ for the coolest dwarfs, as discussed in Cummings et al.\citet{Cummings17}.

Stellar atmosphere models were created from the Kurucz\citet{Kurucz92} models with convective overshoot.
	
\subsection{M48 Metallicity} \label{subsec:feh}

To derive a more robust cluster average metallicity, we have used a subsample of stars that obey the following stringent criteria: a) must be a single (dwarf) member (Section \ref{sec:rv}), b) must have $\sigma_{T_{\rm{eff}}}<$ 75 K (larger $\sigma_{T_{\rm{eff}}}$ may indicate atmospheric problems or other errors leading to unreliable [Fe/H]); and c) {\it v} sin {\it i} $<$ 25 km $s^{-1}$ (the broadened iron lines in stars with larger {\it v} sin {\it i} might be contaminated by nearby lines). We selected 16 non-blended Fe I lines from the solar spectrum\citet[Delbouille et al.][]{Delbouille89} and measured the equivalent width of each line for each star. Fe I lines with an equivalent width greater than 150 m\AA\ were not considered to avoid possible nonlinearity issues. Table \ref{tab:FeI_param} shows the wavelength $\lambda$ (\AA), excitation Potential (eV), and log (gf) values of the 16 Fe I lines. We started with the Kurucz\citet{Kurucz92} atmosphere model grids with [Fe/H] = -0.05 dex, and then used an interpolator to construct model atmospheres using the $T_{\rm{eff}}$, log {\it g}, and $V_{\rm t}$ derived in section \ref{subsec:atmosphere} for each star. Then, we derive A(Fe) by performing local thermal equilibrium (LTE) line analysis for each Fe I line using the {\it abfind} task of MOOG \citet[Sneden et al.][]{Sneden73}. The S/Ns per pixel were measured empirically using the ``line-free'' region from Fe I (6678\AA) and Al I (6696\AA). For stars fainter than $V$ = 14 mag, which all rotate slowly, the ratio of the Poisson-based S/N (from the number of counts) to this empirical S/N is slightly higher than 1, with little scatter from star to star. However, stars with $V\ <$ 14 mag rotate more rapidly, and the ratio deviates from this value increasingly with {\it v} sin {\it i}, possibly because rotational broadening means the line-free region is increasingly less line-free. Table \ref{tab:atmosphere} shows the empirical S/N for $V\ >$ 14 mag, and the Poisson-based S/N divided by this ratio for $V\ <$ 14 mag.

\begin{deluxetable}{ccc}
	\label{tab:FeI_param}
	\tablecaption{Selected Fe I lines}
	\tabletypesize{\scriptsize}
	\tablehead{
		\colhead{Wavelength (\AA)} & \colhead{Excitation Potential (eV)} & \colhead{log (gf)}
	} 
	\startdata
	\hline
	6597.560 & 4.80 & -1.04 \\ 
	6608.044 & 2.28 & -4.02 \\
	6609.118 & 2.56 & -2.67 \\
	6627.540 & 4.55 & -1.57 \\ 
	6653.910 & 4.15 & -2.44 \\
	6677.997 & 2.69 & -1.22 \\
	6703.576 & 2.76 & -3.13 \\
	6710.320 & 1.49 & -4.77 \\ 
	6725.364 & 4.10 & -2.30 \\
	6726.673 & 4.61 & -1.12 \\
	6733.153 & 4.64 & -1.52 \\
	6750.164 & 2.42 & -2.48 \\
	6752.716 & 4.64 & -1.30 \\
	6806.856 & 2.73 & -3.24 \\
	6810.267 & 4.61 & -1.12\\ 
	6820.374 & 4.64 & -1.27 \\
	\hline
	\enddata
\end{deluxetable}

\begin{figure}
	\centering
	\includegraphics[width=0.5\textwidth]{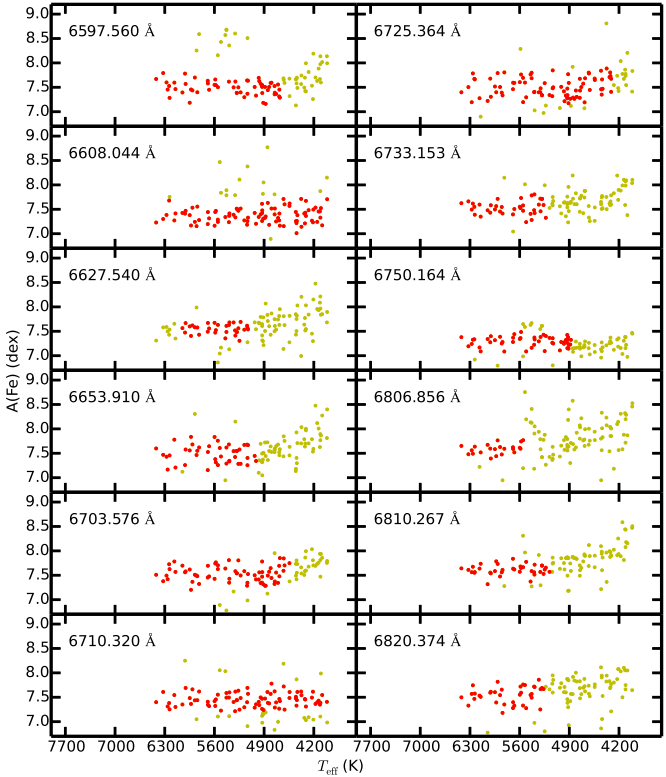}
	\caption{Iron abundance by individual lines. We kept lines shown in red dots. Lines 6609.118\AA, 6677.997 \AA, 6726.673 \AA, and 6752.716 \AA\ are not shown here because the abundances depend on $T_{\rm{eff}}$ throughout the entire range in $T_{\rm{eff}}$. Most lines show an upward trend toward cooler $T_{\rm{eff}}$, and these cooler stars and outliers were eliminated.}
	\label{fig:abun_line}
\end{figure}

\begin{figure}
	\centering
	\includegraphics[width=0.5\textwidth]{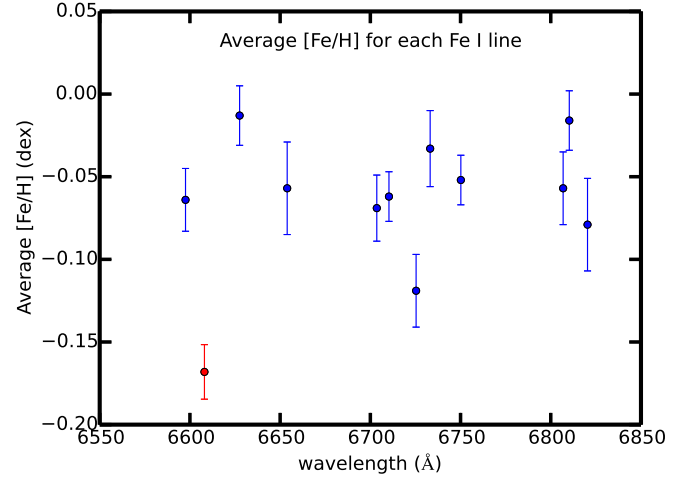}
	\caption{Averaged [Fe/H] over all stars vs. wavelength for each line. Error bars are the standard deviation of the mean. The line at 6609.118 \AA\ was rejected as an outlier.}
	\label{fig:avg_line}
\end{figure}

\begin{figure}
	\centering
	\includegraphics[width=0.5\textwidth]{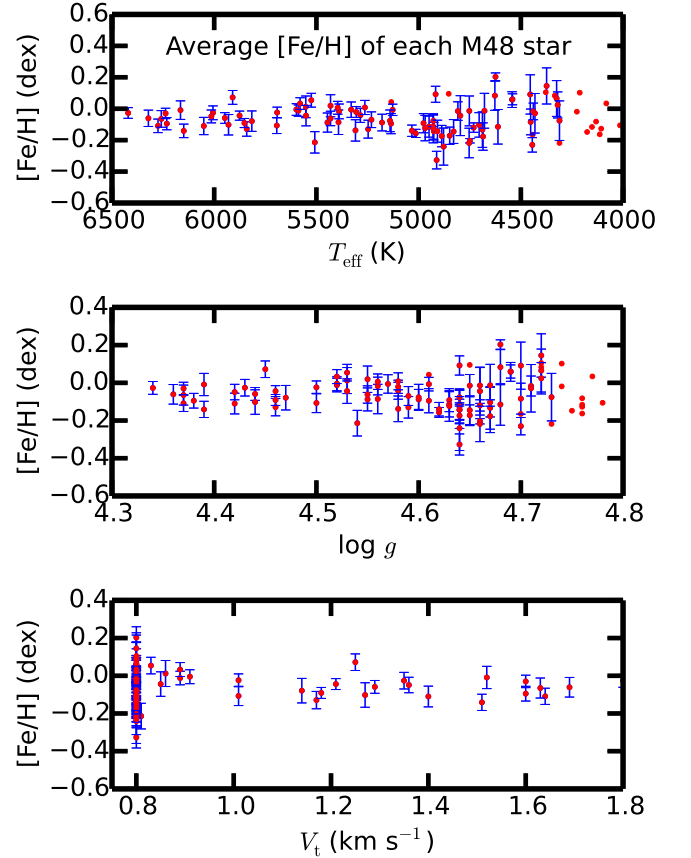}
	\caption{[Fe/H] for individual stars in M48. Error bars are the standard deviation of the mean.}
	\label{fig:avg_star}
\end{figure}

Following Cummings et al.\citet{Cummings17}, Figure \ref{fig:abun_line} shows A(Fe) for each line versus $T_{\rm{eff}}$. Four lines, namely 6609.118 \AA, 6677.997 \AA, 6726.673 \AA, and 6752.716 \AA, not shown in the figure, show trends with $T_{\rm{eff}}$ throughout the entire $T_{\rm{eff}}$ range and were rejected from the calculations of [Fe/H], below. For the remaining 12 Fe I lines (all shown in the figure), we also rejected regions with possible trends in $T_{\rm{eff}}$ and outliers (yellow dots). For the moment, we kept all remaining lines (red dots). Note that most of the lines show a clear upward trend of abundance in the cooler end. This could be due to the effects of spots, deficiencies in the model atmospheres and opacities, unsuspected blends that become important in cooler stars, and other factors; for example, see Schuler et al.\citet{Schuler06, Schuler09} and Maderak et al.\citet{Maderak13}.

To determine [Fe/H] as consistently as possible relative to the Sun, we employed the concept of solar gf-values, as follows. We co-added all of the daytime solar spectra obtained for each configuration to achieve an S/N $\simeq$ 500. Then, by performing the abfind task on the solar spectra, we derived a solar A(Fe) for each selected Fe I line. (We found an overall mean solar A(Fe) = 7.556 $\pm$ 0.0097 dex ($\sigma_{\mu}$), and $\sigma$ = 0.1096 dex.) For each line for each star, we subtracted the solar A(Fe) from the stellar A(Fe) to derive an [Fe/H] for that line for that star. 

Figure \ref{fig:avg_line} shows A(Fe) for each line as averaged over all (kept) stars, plotted against wavelength. This illustrates that the average abundances are consistent from line to line, except for the 6608.044 \AA\ line (red), which we rejected as a 2.6$\sigma$ outlier and and which we excluded from further analysis. For all surviving lines, we subtracted the solar A(Fe) from that line's A(Fe) to derive that line's [Fe/H]. A linear (not log) average of each star's lines produced a [Fe/H] for that star (as in Boesgaard et al.\citet{Boesgaard05} and Cummings et al.\citet{Cummings17}). Figure \ref{fig:avg_star} shows the stellar [Fe/H] versus $T_{\rm{eff}}$ (top panel), [Fe/H] versus log {\it g} (middle panel), and [Fe/H] versus $V_{\rm t}$ (bottom panel) for all stars in our carefully selected subsample. {\it Across nearly the entire range in $T_{\rm{eff}}$ of 2500 K, [Fe/H] shows no dependence on $T_{\rm{eff}}$}. Similarly, no trends are found with log {\it g} or $V_{\rm t}$. Table \ref{tab:atmosphere} also lists the stellar [Fe/H] for those stars included in the Fe-subsample, and their errors (standard deviation of the mean). 

The overall cluster average [Fe/H] for M48 was determined using the precepts discussed in Cummings et al.\citet{Cummings17} and our other works. In particular, the cluster [Fe/H] was derived by averaging linearly and in linear (not log) space all the lines surviving the cuts in Figures \ref{fig:abun_line} and \ref{fig:avg_line}. The result is [Fe/H]$_{\rm{M48}}$ = -0.063 $\pm$ 0.007 dex ($\sigma_{\mu}$, and $\sigma$ = 0.151 dex), in excellent agreement with our photometric study. Table \ref{tab:sys_error} shows how systematic changes in $\Delta(E(\bv))$, $\Delta$(log {\it g}), and $\Delta(V_{\rm t})$ affect the derived [Fe/H].

The only spectroscopic abundance (that we could find) comes from Wallerstein \& Conti\citet{Wallerstein64}, who observed one giant and report [Fe/H] = -0.51 dex. But they also indicate that the star is metal poor by a factor of two compared to $\gamma\ Tau$ of the Hyades, which suggests [Fe/H] = -0.15 dex, assuming [Fe/H] = +0.15 dex for the Hyades\citet[Cummings et al.][]{Cummings17}. They quote an uncertainty in [Fe/H] by a factor of two, so their result is in agreement with ours. There are several photometrically based metallicities. From Claria's\citet{Claria85} DDO photometry of three giants, we infer [Fe/H] = +0.14 $\pm$ 0.05 dex (using their equation (1) and section 5.10). The Claria [Fe/H] is quoted as [Fe/H] = +0.04 dex in Strobel\citet{Strobel91} and has been recalibrated to [Fe/H] = +0.01 $\pm$ 0.02 dex in Piatti et al.\citet{Piatti95} and to +0.08 $\pm$ 0.014 dex in Twarog et al.\citet{Twarog97}. Strobel\citet{Strobel89} lists [Fe/H] = -0.02 dex, which is recalibrated to [Fe/H] = +0.03 dex in Strobel\citet{Strobel91}. Hog \& Flynn\citet{Hog98} list [Fe/H] = -0.13 dex. The next three photometric studies employ numerous stars. Rider et al.\citet[$u'g'r'i'z'$]{Rider04} report a range of [Fe/H] = -0.1 to +0.1 dex, with a preferred value of 0.0 dex. Wu et al.\citet[BATC]{Wu05} report [Fe/H] = 0.0 dex (no error). Finally, Balaguer-Nunez et al.\citet[$uvby-H_{\beta}$]{Balaguer05} report [Fe/H] = -0.24 $\pm$ 0.27 dex.

\begin{deluxetable}{ccccc}
	\label{tab:sys_error}
	\tablecaption{Possible systematic errors on [Fe/H] (dex)}
	\tabletypesize{\tiny}
	\tablehead{
		\colhead{Parameter Changes$^1$} & \colhead{4300 K}$^2$ & \colhead{5300 K}$^2$ & \colhead{6300 K}$^2$ & M48 cluster$^3$
	} 
	\startdata
	\hline
	$\Delta(E(\bv))$ = +0.01 mag& -0.011 & 0.017 & 0.027 & 0.0089 \\ 
	$\Delta$(log {\it g}) = +0.2 & 0.033 & -0.005 & -0.005 & 0.0076 \\
	$\Delta(V_{\rm t})$ = +0.2 km s$^{-1}$& -0.017 & -0.028 & -0.016 & -0.023 \\
	$\Delta_{comb}(\Delta(E(\bv$)) = +0.01 mag) & -0.012 & 0.018 & 0.023 & 0.0012 \\ 
	\hline
	\enddata
	\tablecomments{1. The first three lines show changes for each of the three parameters (E($\bv$), log {\it g}, $V_{\rm t}$) independently. For example, we change $E(\bv$) by +0.01 mag, but keep the log {\it g} and $V_{\rm t}$ the same. The bottom line shows changes for all three parameters simultaneously based on $\Delta(E(\bv$)) = +0.01 mag: $\Delta(E(\bv$)) implies a certain $\Delta$(log {\it g}), and these two imply a certain $\Delta(V_{\rm t})$.
		2. Change of [Fe/H] for a star at $T_{\rm{eff}}$ = 4300 K, 5300 K, 6300 K.
		3. Change of [Fe/H] for the whole M48 cluster following the above procedure.}
\end{deluxetable}

\section{Summary}

We present high signal-to-noise WIYN/Hydra spectra for 287 stars that mainly fall on the single-star fiducial main-sequence of our M48 CMD. We report radial velocities ($V_{\rm{RAD}}$) for all of the stars on at least two nights (except one possible red-giant star member, which was observed only once) and compare the $V_{\rm{RAD}}$ from different nights along with the Fourier-transformed spectra to determine binarity for all of the stars. Using only single stars with rotational velocity ({\it v} sin {\it i}) less than 20 km s$^{-1}$ and $\sigma(V_{\rm{RAD}}) <$ 1 km s$^{-1}$, we derive an initial estimate for the M48 cluster mean $V_{\rm{RAD}}$ of 8.399 $\pm$ 0.037 km s$^{-1}$ ($\sigma_{\mu}$, and $\sigma$ = 0.099 km s$^{-1}$). Stars within $2\sigma$ of the M48 $V_{\rm{RAD}}$ are defined as radial velocity members. We retrieve the proper motion in R.A. and decl. and parallax of stars from the Gaia DR2\citet[Gaia Collaboration][]{Gaia16, Gaia18} to determine independently a group of highly probable M48 members. Combining both the $V_{\rm{RAD}}$ data and the Gaia DR2 data, we designate 152 stars as single members of M48 (sm), 11 stars as binary members (bm), 16 stars as members of uncertain multiplicity (?m), 56 stars as single-star nonmembers (sn), 28 as single-star ``likely" nonmembers (sn?), two as single-star ``likely" members (sm?), one star as a binary ``likely" member (bm?), five stars as binary nonmembers (bn), 10 stars as ``likely" members of uncertain multiplicity (?m?), three stars as nonmembers of uncertain multiplicity (?n), and three stars as ``likely" nonmembers of uncertain multiplicity (?n?). Now, using a more restricted sample of stars, namely, (1) it must be sm, (2) {\it v} sin {\it i} $<$ 20 km s$^{-1}$, and (3) $\sigma_{V_{\rm{RAD}}} <$ 1.0 km s$^{-1}$, we evaluate our final M48 cluster mean $V_{\rm{RAD}}$ as 8.512 $\pm$ 0.087 km s$^{-1}$ ($\sigma_{\mu}$).

Using our spectroscopic data together with Gaia DR2 data, we find a minimum binary fraction in M48 of 11--21\%. This is similar to a number of other clusters that span a variety of ages and richness classes but not as high as some, such as the Hyades.

To derive a more robust cluster average metallicity, we use a subsample of stars that obey the following stringent criteria: must be a single (dwarf) member, must have $\sigma_{T_{\rm{eff}}}$ derived from 10 color index combinations of $UBVRI$ photometry $<$ 75 K, and {\it v} sin {\it i} $<$ 25 km s$^{-1}$. Stellar parameters are evaluated as follows. We use the averaged $\bv$ color transformed from all 10 possible color combinations of $UBVRI$ to determine the effective temperature ($T_{\rm{eff}}$) for each star from our usual color--metallicity--temperature relation\citet[Cummings et al.][]{Cummings17}. The log {\it g} values are derived from $Y^2$ isochrones\citet[Demarque et al.][]{Demarque04} based on a cluster age of 420 Myr and [Fe/H] = -0.05 dex, and we adopt the relationship from Edvardsson et al.\citet{Edvardsson93} to determine $V_{\rm t}$, with a lower limit of 0.8 km s$^{-1}$ for the coolest stars. Using the Kurucz\citet{Kurucz92} stellar atmospheres, we derive A(Fe) for all Fe I lines. Solar A(Fe) are calculated based on our high--S/N daytime sky spectra in a similar way and are then subtracted to arrive at [Fe/H] for each line in each star. Examining each Fe I line separately for all stars as a function of $T_{\rm{eff}}$, we eliminate portions (or entire lines) that show trends with $T_{\rm{eff}}$, and outliers. The average values of A(Fe) from each line are consistent with each other, except for the 6608.044 \AA\ line, which we reject as a 2.6$\sigma$ outlier. We use surviving lines to derive average (in linear space) [Fe/H] for individual stars (Table \ref{tab:atmosphere}), and we average the entire set of surviving lines with linear weighting and in linear space to determine the cluster average M48, which gives [Fe/H] = -0.063 $\pm$ 0.007 dex ($\sigma_{\mu}$, and $\sigma$ = 0.151 dex). The stellar [Fe/H] show no trend with $T_{\rm{eff}}$ over an unprecedentedly large range of 2500 K in $T_{\rm{eff}}$, increasing over the range of 1700 K used by Cummings et al.\citet{Cummings17} for Praesepe.

While the metallicities of open clusters drop as a function of Galactocentric distance from near-solar in the solar neighborhood to subsolar toward the periphery (Twarog et al.\citet{Twarog97}; Jacobson et al.\citet{Jacobson11}), those within approximately 1 kpc of the solar neighborhood that have been measured precisely span at most a range of about [Fe/H] = 0.3 dex and show no relation to age (Boesgaard\citet{Boesgaard89}, Friel \& Boesgaard\citet{Friel92}). These authors indicate that their errors are sufficiently small so that the cluster metallicities are distinguishable, and thus, the absence of a metallicity--age relation is not merely a reflection of scatter. The clusters considered span a range in ages from 50 Myr to 4 Gyr. Boesgaard\citet{Boesgaard89} and Friel \& Boesgaard\citet{Friel92} conclude that the gas from which these clusters formed preserved small but significant differences in [Fe/H] and that the mixing time scale in the solar neighborhood is at least several billion years. Our metallicity for M48 is consistent with these conclusions, for example, by being more metal poor than both younger objects (Pleiades) and older objects (Hyades, Praesepe).

\bigskip

C.P.D. acknowledges support from the NSF through grant AST-1909456. This research has made use of the WEBDA database, operated at the Department of Theoretical Physics and Astrophysics of the Masaryk University. We would also like to thank the WIYN observatory staff, whose dedication and skillful work helped us obtain these excellent spectra. This work has made use of data from the European Space Agency (ESA) mission Gaia (\url{https://www.cosmos.esa.int/gaia}), processed by the Gaia Data Processing and Analysis Consortium (DPAC, \url{https://www.cosmos.esa.int/web/gaia/dpac/consortium}). Funding for the DPAC has been provided by national institutions, in particular, the institutions participating in the Gaia Multilateral Agreement.

\bibliographystyle{aasjournal}

\end{document}